\documentclass[twocolumn,a4paper,reprint,amssymb,aps,prd,groupedaddress,superscriptaddress,nofootinbib]{revtex4-2}
\pdfoutput=1
\usepackage{booktabs}
\usepackage{multirow}
\usepackage{physics}
\usepackage{dcolumn}
\usepackage{bm}
\usepackage{amsmath}
\usepackage{mathtools}
\usepackage{amsfonts}
\usepackage{pbox}
\usepackage{color}
\usepackage[dvipsnames]{xcolor}
\usepackage{tabularx}
\usepackage{subfigure}
\usepackage{graphicx}
\usepackage{psfrag}
\usepackage{lipsum}
\usepackage{enumitem}
\usepackage{changes}
\usepackage{array}
\usepackage{CJK}
\usepackage[english]{babel}

\usepackage[colorlinks = true,
            linkcolor = blue,
            urlcolor  = blue,
            citecolor = blue,
            anchorcolor = blue]{hyperref}
\usepackage{blindtext}
\usepackage[utf8]{inputenc}
\usepackage{graphicx}
\usepackage{placeins}
\usepackage{siunitx}
\usepackage{amssymb}
\usepackage{mathtools}
\usepackage{physics}
\usepackage{tikz}
\usepackage[titletoc]{appendix}
\usepackage{listings}
\usepackage{float}
\usepackage{braket}

\usepackage[capitalize]{cleveref}
\usepackage{soul}
\usepackage{array}

\begin{document}

\title{Measuring Hubble constant in an anisotropic extension of $\Lambda$CDM model}
\author{Vikrant Yadav}
\email{vikuyd@gmail.com}
\affiliation{School of Basic and Applied Sciences, Raffles University, Neemrana - 301705, Rajasthan, India}

\begin{abstract}
Cosmic Microwave Background (CMB) independent approaches are frequently used in the literature to provide estimates of Hubble constant ($H_0$). In this work, we report CMB independent constraints on $H_0$ in an anisotropic extension of $\Lambda$CDM model using the Big Bang Nucleosynthesis (BBN), Baryonic Acoustic Oscillations (BAOs), Cosmic Chronometer (CC), and Pantheon+ (PP) compilation of Type Ia supernovae and SH0ES Cepheid host distance anchors data. In the anisotropic model, we find $H_{\rm 0}=70.1^{+1.2}_{-1.5}\; (72.67\pm 0.85)\;\rm km\, s^{-1}\, Mpc^{-1}$ both with 68\% CL from BAO+BBN+CC+PP (BAO+BBN+CC+PPSH0ES) data.  The analyses of the anisotropic model with the two combinations of data sets reveal that anisotropy is positively correlated with $H_0$, and an anisotropy of the order $10^{-14}$ in the anisotropic model reduces the $H_0$ tension by $\sim 2\sigma$.
\end{abstract}


\maketitle

\section{Introduction}


\label{sec:intro}
The scientific community is enthralled by the revelation of the Universe's accelerated expansion~\cite{SupernovaSearchTeam:1998fmf,SupernovaCosmologyProject:1998vns}. Among the various approaches to explain the accelerated expansion of the Universe, the most straightforward and well-established one is to include an exotic fluid with negative pressure, dubbed as dark energy (DE), within the framework of Einstein's theory of general relativity (GR), primarily limited to the Friedmann-Lemaitre-Robertson-Walker (FLRW) class of solutions, which fuels the Universe's late-time accelerated expansion~\cite{Copeland:2006wr,Bamba:2012cp}. Various DE cosmological models have been proposed in the literature to explain the dynamics of the Universe; however, the search for the actual cosmological model capable of accurately explaining the accelerated expansion of the universe is still ongoing. The $\Lambda$CDM cosmological model, the mathematically simplest model with just two heavy ingredients, namely, the positive cosmological constant ($\Lambda >0$) mimicking DE and the cold dark matter (CDM), is considered as the most successful model introduced so far in the literature as it finds consistency with observational data from various astrophysical and cosmological probes,  including the recent measurement of the cosmic microwave background (CMB) temperature at high redshift from H$_2$O absorption~\cite{Riechers:2022tcj}. As a result of methodological advancement and the decrease of errors on the calculated cosmological parameters, some statistically significant tensions in the measurement of various quantities from the  CMB data and late-time cosmological model independent probes have emerged. The scale and consistency of these inconsistencies across probes clearly suggest that the classic cosmic scenario may have failed and that new physics is required, even though some of these discrepancies may ultimately be attributable to systematic errors in the experiments \cite{Abdalla:2022yfr}. 

The present-day expansion rate of the Universe is given by the Hubble constant $H_0$. The most statistically significant and persistent discrepancy, known as the Hubble tension, appears between the estimated values of Hubble constant $H_0$ from the CMB data within vanilla $\Lambda$CDM cosmology, and the direct local distance ladder measurements. In particular, the Hubble tension is referred to as the disagreement at $5.0\sigma$ between the latest SH0ES  (Supernovae and $H_0$ for the Equation of State of dark energy) collaboration \cite{Riess:2021jrx} constraint, $H_0^{\rm R22}=(73.04 \pm 1.04){\rm \,km\,s^{-1}\,Mpc^{-1}}$ at 68\% confidence level (CL), based on the supernovae calibrated by Cepheids, and the {\it Planck} collaboration~\cite{Planck:2018vyg} value, $H_0=\left(67.27\pm0.60\right){\rm \,km\,s^{-1}\,Mpc^{-1}}$ at 68\%  CL. The possibility of the Hubble tension calling for new physics, potentially related to the dark sector, is now taken very seriously, with a wide range of proposals having been made \cite{DiValentino:2021izs,Abdalla:2022yfr}. Modifications to CDM are considered either before recombination or in the late Universe. It is now widely accepted that late-time models are less successful in resolving the Hubble tension because, as $H_0$ is raised, their fit to Baryonic Acoustic Oscillation (BAO) and Hubble flow Type Ia supernovae (SNe Ia) data become poorer because the sound horizon at baryon drag $r_{\rm{d}}$ is unaffected. The emphasis is currently mostly on early-time adjustments, which seek to decrease $r_{\rm{d}}$ in order to obtain a larger $H_0$ while maintaining the angular size of the sound horizon to the CMB value and avoiding late-time BAO and SNe Ia limitations. Examples of such models include, but are not limited to, those that alter the recombination history or increase the pre-recombination growth rate \cite{Arendse:2019hev,Cai:2021weh}. Resolution of the $H_0$ tension is also proposed by introducing a physical phenomenon at intermediate redshifts, for instance, a rapid transition of the Universe from anti-de Sitter vacua to de Sitter vacua (viz., the cosmological constant switches sign from negative to positive) at redshift $z\sim 2$, helps to alleviate the $H_0$ tension  \cite{Akarsu:2019hmw,Akarsu:2021fol,Akarsu:2022typ,Akarsu:2023mfb}. See \cite{DiValentino:2021izs,Abdalla:2022yfr,Perivolaropoulos:2021jda,Hu:2023jqc,Kroupa:2023ubo,Vagnozzi:2023nrq} for the current status and various approaches/models for the resolution of $H_0$ tension.

Cosmological Principle (CP), the idea that the Universe is spatially isotropic and homogenous on vast scales, is what has driven cosmology forward for the past century. It is typically expressed using the FLRW cosmologies as the spacetime metric, which leads to the development of the successful and highly predicative $\Lambda$CDM model. As discussed above. the statistically significant disparity in the Hubble constant, $H_0$, value is the most prominent example of the tensions that have developed inside the $\Lambda$CDM model. Since the CP is closely related to the idea of cosmic expansion regulated by a single parameter, the consequences of the $H_0$ tension may go beyond the $\Lambda$CDM to the CP itself. Our universe may be anisotropic and introduces possible observational evidence in Type Ia supernovae (SNe Ia), Gamma Ray Bursts (GRBs), Quasars, and Galaxies (see \cite{Hu:2020mzd} and references therein). The review \cite{Aluri:2022hzs} analyses current observational signatures for deviations from the CP, identifying the synergies and disagreements that motivate for further research in this direction.  

The simplest anisotropic generalization of the standard $\Lambda$CDM model is investigated in \cite{Akarsu:2019pwn}, where the spatially flat RW metric is repalced by the Bianchi type-I metric while keeping the physical ingredients of the Universe in the model as usual. This modification brings a new term $\Omega_{\sigma 0}a^{-6}$, where $\Omega_{\sigma 0}$ is the present-day expansion anisotropy parameter and $a$ is the average expansion scale factor, to the Friedmann equation of the standard $\Lambda$CDM model, by rendering $H(a)$ as the average expansion rate of the Universe. The authors obtained  tight constraints on $\Omega_{\sigma0}$, viz., $\Omega_{\sigma0}\lesssim10^{-15}$  in the analyses of the anisotropic model with the Baryonic Acoustic Oscillations (BAO) and CMB data where the drag redshift as $z_{\rm d}=1059.6$ (involved in BAO analysis) and the last scattering redshift as $z_* = 1089.9$ (involved in CMB data) were fixed in order to guarantee expansion anisotropy to remain as correction all the way to the largest redshifts. A similar study was carried out in \cite{Akarsu:2021max} where anisotropic expansion and curvature were studied together. Anisotropic models are constrained with observational data in different contexts, see \cite{Akarsu:2019pwn,Akarsu:2021max,Amirhashchi:2018bic,Amirhashchi:2018nxl,Amirhashchi:2020qep} and references therein. 

CMB independent approaches are frequently used in the literature to provide estimates of $H_0$ \cite{Schoneberg:2019wmt,Schoneberg:2022ggi}. In our present study, we investigate the anisotropic model considered in \cite{Akarsu:2019pwn} with the aim to obtain CMB-independent constraints on $H_0$ within the framework of anisotropic model, and also to study the effect of anisotropy on $H_0$. The paper is organized as follows: In Section \ref{sec2}, we describe the governing equations and free parameters of the anisotropic model. In Section \ref{sec3}, we present the data-sets and methodology used for the anaysis of the anisotropic model. In Section \ref{sec4}, we report the observational constraints on the model parameters and discuss the results. We conclude main findings in Section \ref{sec5}.

\section{The Anisotropic Model}\label{sec2}
We consider Bianchi type I metric, which carries different scale factors in three orthogonal directions, and is given by
\begin{equation}\label{metric}
    \text{d}s^2=-\text{d}t^2 + A^2\text{d}x^2 + B^2\text{d}y^2 + C^2\text{d}z^2,
\end{equation} 
where $\{A,B,C\}$ are the directional scale factors along the principal axes $\{x,y,z\}$ and are functions of cosmic time $t$ only. We define the average expansion scale factor: $a=(ABC)^{\frac{1}{3}}$, and the average Hubble parameter:  $H=\frac{\dot{a}}{a}=\frac{1}{3}\left(H_x + H_y+H_z\right)$, where $H_x =\frac{\dot{A}}{A}$, $H_y =\frac{\dot{B}}{B}$, and $H_z =\frac{\dot{C}}{C}$ are the directional Hubble parameters defined along the principal axes $\{x,y,z\}$.

The Einstein field equations of GR are given by
\begin{equation}\label{fieldeqn}   
G_{\mu\nu}\equiv R_{\mu\nu}-\frac{1}{2}g_{\mu\nu} R= 8\pi GT_{\mu\nu},
\end{equation}    
where, on the left hand side, $G_{\mu\nu}$ is the Einstein tensor, $R_{\mu\nu}$ is the Ricci tensor, $R$ is the Ricci scalar, and $g_{\mu\nu}$ is the metric tensor. On right hand side, $G$ is Newton's gravitational constant, and $T_{\mu\nu}$ is the energy momentum tensor, which for a perfect fluid with energy density $\rho$ and pressure $p$ is of the form 
\begin{equation}\label{emt}
 T^{\nu}_{\,\,\mu} = \text{diag} [-\rho, p, p, p].    
\end{equation}
As a consequence of the twice-contracted Bianchi identity, that is, $G^{\mu\nu}_{\;\;\;;\nu}=0$, the Einstein field equations ~\eqref{fieldeqn} satisfy the conservation equation
\begin{equation}\label{ece}
 T^{\mu\nu}_{\;\;\;;\nu}=0.    
\end{equation}
For the perfect fluid matter distribution, it simplifies to 
\begin{equation}\label{ce}
 \dot{\rho}+3H(\rho+p)=0,    
\end{equation}
where the dot represents the derivative with respect to the time $t$. For a perfect fluid $i$, with energy density $\rho_i$, pressure $p_i$ and constant equation of state (EoS) $w_i=p_i/\rho_i$, the evolution of its energy density $\rho_i$ is given by the continuity equation \eqref{ce} as follows: 
\begin{equation}\label{eos}
\rho_i=\rho_{i0}a^{-3(1+w_i)},    
\end{equation}
where $\rho_{i0}$ is the present-day value of $\rho_{i}$, that is at $a=a_0=1$, being the present-day value of $a$. Here and onward, a subscript 0 attached to any quantity would mean its value in the present-day Universe. 
We assume that the Universe is filled with the usual cosmological fluids, namely the radiation (photons and neutrinos) with EoS $w_{\rm r}=p_{\rm r}/\rho_{\rm r}=\frac{1}{3}$, pressureless fluid (baryonic and cold dark matter) with EoS $w_{\rm m}=p_{\rm m}/\rho_{\rm m}=0$, and dark energy mimicked by cosmological constant $\Lambda$ with EoS $w_{\Lambda}=p_{\Lambda}/\rho_{\Lambda}=-1$. 
We assume these energy sources interact only gravitationally, so that the continuity equation \eqref{ce} is satisfied separately by each source, and in view of \eqref{eos}, this leads to
\begin{equation}\label{sources}
\rho\equiv\rho_\text{r}+\rho_\text{m}+\rho_{\Lambda}= \rho_{\text{r}0}a^{-4}+\rho_{\text{m}0} a^{-3}+\rho_{\Lambda}
\end{equation}

Now, the Einstein field equations~\eqref{fieldeqn} for the Bianchi type I metric~\eqref{metric} lead to the following set of differential equations:
\begin{align}
    \frac{\dot{A}}{A}\frac{\dot{B}}{B}+ \frac{\dot{B}}{B}\frac{\dot{C}}{C}+ \frac{\dot{A}}{A}\frac{\dot{C}}{C}&=8\pi G \rho \label{BI.1}, \\ 
    -\frac{\Ddot{B}}{B}-\frac{\Ddot{C}}{C}- \frac{\dot{B}}{B}\frac{\dot{C}}{C}&=8\pi G p \label{BI.2}, \\ 
      -\frac{\Ddot{A}}{A}-\frac{\Ddot{C}}{C}- \frac{\dot{A}}{A}\frac{\dot{C}}{C}&=8\pi G p \label{BI.3} , \\
        -\frac{\Ddot{A}}{A}-\frac{\Ddot{B}}{B}- \frac{\dot{A}}{A}\frac{\dot{B}}{B}&=8\pi G p \label{BI.4} . 
\end{align}
The shear scalar, in terms of the directional Hubble parameters, is given by
\begin{equation}\label{BIshear} 
\sigma^2 = \frac{(H_x -H_y)^2 + (H_y - H_z)^2 + (H_z -H_x)^2}{6}.
\end{equation} 
The equations~\eqref{BI.1}-\eqref{BI.4} can be recast as follows: 
\begin{align} 
3H^2-\sigma^2=8\pi G\rho,\label{hubble0}\\
     \dot{H}_x-\dot{H}_y+3H(H_x-H_y)=0,\label{hubble1} \\
      \dot{H}_y-\dot{H}_z+3H(H_y-H_z)=0,\label{hubble2} \\  
       \dot{H}_z-\dot{H}_x+3H(H_z-H_x)=0\label{hubble3}.    
\end{align} 
Using the time derivative of $\sigma^2$ given in Eq.~\eqref{BIshear} along with these equations \eqref{hubble1}-\eqref{hubble3}, we obtain the following shear propagation equation:
\begin{equation}\label{BIshearp}     
    \dot{\sigma}+3H\sigma=0, 
\end{equation}     
which on integration leads to
\begin{equation}\label{BIsigma2}  
\sigma^2 = \sigma_0^2 a^{-6}.
\end{equation}
Using equations \eqref{sources} and \eqref{BIsigma2} into \eqref{hubble0}, we obtain
\begin{equation}\label{themodel} 
    \frac{H^2}{H_0^2}  =\Omega_{\sigma0}a^{-6}+\Omega_{\text{r}0}a^{-4}+\Omega_{\text{m}0}a^{-3}+\Omega_{\Lambda0},   
\end{equation}  
where $\Omega_{\sigma 0}=\frac{\sigma_0^2 }{3H_0^2}$ is the expansion anisotropy parameter, $\Omega_{\rm r0}=\frac{8\pi G}{3H_0^2}\rho_{\rm r0}$ is the radiation density parameter, $\Omega_{\rm m0}=\frac{8\pi G}{3H_0^2}\rho_{\rm m0}$ is the matter density parameter, and $\Omega_{\Lambda 0}=\frac{8\pi G}{3H_0^2}\rho_{\rm \Lambda}$ is the DE density parameter, satisfying $\Omega_{\sigma0} + \Omega_{\text{r}0}+ \Omega_{\text{m}0}+\Omega_{\Lambda0}=1$. Note that here expansion anisotropy parameter $\Omega_{\sigma0}$ is a non-negative geometric term. Further, the absolute CMB monopole temperature measured by Far Infrared Absolute Spectrophotometer (FIRAS), $T_0 =2.7255 \pm 0.0006 \,{\rm K}$~\cite{Fixsen:2009ug}, extremely well constrains the present-day radiation energy density, viz., $\Omega_{\rm r0}\equiv \rho_{\rm r0}/3H_0^2=2.469\times10^{-5}h^{-2}(1+0.2271N_{\rm eff})$, where $h=H_0/100\, {\rm km\,s}^{-1}{\rm Mpc}^{-1}$ is the dimensionless reduced Hubble constant and $N_{\rm eff}=3.046$ is the standard number of effective neutrino species with minimum allowed mass $m_{\nu}=0.06$~eV.

Notice that equation \eqref{themodel} is the generalized Friedmann equation, describing the spatially flat and homogeneous but not necessarily isotropic extension of the $\Lambda$CDM model. In what follows, we will denote this anisotropic model by $\Lambda$CDM+$\Omega_{\sigma0}$.

\section{Data and methodology}\label{sec3}
\noindent\textbf{BAO}: Recent BAO measurements employing galaxies, Lyman-$\alpha$ (Ly$\alpha$) and quasars,  are presented for the completed experiments by the Sloan Digital Sky Survey (SDSS)~\cite{eBOSS:2020yzd}. A compilation of data from BOSS, eBOSS, SDSS, and  SDSS-II  is included in these experiments, which makes independent BAO measurements of Hubble distances and angular-diameter distances relative to the sound horizon from eight different samples available, as indicated in Table~\ref{tab:BAO_measurements}.

\begin{table}[b!]
\caption{\rm Clustering measurements for the BAOs  from Ref.~\cite{eBOSS:2020yzd}.}
 \scalebox{0.9}{
\centering
\begin{tabular}{l|c|c|c|c}
\toprule
\hline
\textbf{Parameter} & $\bm{\,\,z_{\rm eff}\,\,}$  &  $\bm{\,\,D_V(z)/r_{\rm d}\,\,}$ & $\bm{\,\,D_M(z)/r_{\rm d}\,\,}$ & $\bm{\,\,D_H(z)/r_{\rm d}\,\,}$  \\
\hline

\hline
MGS & 0.15 & $4.47 \pm 0.17$ & --- & --- \\

BOSS Galaxy & $0.38$ & --- & $10.23 \pm 0.17$ & $25.00 \pm 0.76$ \\

BOSS Galaxy & $0.51$ & --- & $13.36 \pm 0.21$ & $22.33 \pm 0.58$ \\

eBOSS LRG & $0.70$ & --- & $17.86 \pm 0.33$ & $19.33 \pm 0.53$ \\

eBOSS ELG & $0.85$ & $18.33_{-0.62}^{+0.57}$ & --- & --- \\

eBOSS Quasar & $1.48$ & --- & $30.69 \pm 0.80$ & $13.26 \pm 0.55$ \\

Ly$\alpha$-Ly$\alpha$ & $2.33$ & --- & $37.6 \pm 1.9$ & $8.93 \pm 0.28$ \\

Ly$\alpha$-Quasar & $2.33$ & --- &$37.3 \pm 1.7$ & $9.08 \pm 0.34$ \\
\hline
\bottomrule
\hline
\end{tabular} }
\label{tab:BAO_measurements}
\end{table}

The comoving size of the sound horizon ($r_{\rm s}$) at the drag redshift ($z_{\rm d}$), i.e., $r_{\rm d}$, is given by
\begin{equation}{
\label{comoving size}
r_{\rm d}=r_{\rm s}(z_{\rm d})=\int_{z_{\rm d}}^\infty \frac{c_{\rm s}\text{d}z}{H(z)},} 
\end{equation}
where $c_{\rm s}=\frac{c}{\sqrt{3(1+\mathcal{R})}}$ is the sound speed of the baryon--photon fluid; $\mathcal{R}=\frac{3\Omega_{\rm b0}}{4\Omega_{\rm \gamma 0}(1+z)}$ with $\Omega_{\rm b0}=0.022h^{-2}$ and $\Omega_{\gamma 0}=2.469\times 10^{-5}h^{-2}$ respectively being the present-day physical density of baryons and  the present-day physical density of photons ~\cite{Cooke:2016rky,Bennett:2020zkv}.

The quantities $D_H(z)/r_{\rm d}$ and $D_{M}(z)/r_{\rm d}$ are directly constrained by the BAO measurements. The Hubble distance is given by
\begin{equation}
 D_H(z) = \frac{c}{H(z)}.
\end{equation}
The comoving angular diameter distance $D_{M}(z)$ for flat cosmology is calculated as
\begin{equation}
    D_{M}(z)= {c \over H_0}\int_0^z \text{d}z' {H_0 \over H(z')}.
\label{eqn:dcomove}
\end{equation}

The spherically averaged distance $D_V(z)$ is defined as
\begin{equation}
 D_V(z) \equiv \left[z D^2_M(z) D_H(z)\right]^{1/3}.
\end{equation}

Denoting $d_1$ for $D_{V}(z)/r_{\rm d}$, $d_2$ for $D_{M}(z)/r_{\rm d}$, and $d_3$ for $D_{H}(z)/r_{\rm d}$, we define the chi-squared function for each measurement in Table~\ref{tab:BAO_measurements} as follows: 
\begin{eqnarray}
\chi^2_{\rm B_1} =&\displaystyle\sum_{i=1}^{2}\left(\frac{d_1^{\text{obs}}(z_i)-d_1^{\text{th}}(z_i)}{\sigma_{d_1^{\text{obs}}(z_i)}}\right)^2,\nonumber \\
\chi^2_{\rm B_2}=&\displaystyle\sum_{j=1}^{6}\left(\frac{d_2^{\text{obs}}(z_j)-d_2^{\text{th}}(z_j)}{\sigma_{d_2^{\text{obs}}(z_j)}}\right)^2,\nonumber \\
\chi^2_{\rm B_3}=&\displaystyle\sum_{j=1}^{6}\left(\frac{d_3^{\text{obs}}(z_j)-d_3^{\text{th}}(z_j)}{\sigma_{d_3^{\text{obs}}(z_j)}}\right)^2.
\end{eqnarray}

Here, $d^{\rm obs}$ stands for the observed distance value as presented in Table~\ref{tab:BAO_measurements}, while $d^{\rm th}$ represents the calculated theoretical value for the models being examined. Concerning the $D_{V}(z)/r_{\rm d}$ scenario, the effective redshifts for the two samples, MGS and eBOSS ELG, are denoted as $z_i$ ($i=1,2$). As for the $D_{M}(z)/r_{\rm d}$ and $D_{H}(z)/r_{\rm d}$ cases, the effective redshifts for the six measurements are respectively labeled as $z_j$ ($j=1,2,3,4,5,6$), corresponding to BOSS Galaxy, BOSS Galaxy, eBOSS Galaxy, eBOSS LRG, eBOSS Quasar, Ly$\alpha$-Ly$\alpha$, and Ly$\alpha$-Quasar.

Therefore, the cumulative chi-squared expression for the BAO measurements, represented as $\chi^2_{\rm BAO}$, is formulated as follows:
\begin{equation}
\chi^2_{\rm BAO}=\chi^2_{\rm B_1} +\chi^2_{\rm B_2}+\chi^2_{\rm B_3}.
\end{equation}


\noindent\textbf{CC}: The Cosmic Chronometer (CC) employs a robust method to track the historical expansion of the universe by analyzing $H(z)$ measurements. A comprehensive collection of 33 $H(z)$ measurements from CC has been assembled, encompassing redshift values ranging from 0.07 to 1.965. These measurements are meticulously detailed in Table \ref{tab:CC} along with their respective references. The fundamental concept underpinning these measurements was originally introduced in \cite{Jimenez:2001gg}, establishing a fundamental link connecting the Hubble parameter $H(z)$, redshift $z$, and cosmic time $t$:
\begin{equation}
    H(z)= -\frac{1}{1+z}\frac{\text{d}z}{\text{d}t}.
\end{equation}

\begin{table}[t!]
\caption{\rm Compilation of CC measurements of $H(z)$.}

\begin{center}
\begin{tabular}{lllr}
\multicolumn{4}{c}{{}}\\
\hline \toprule\hline
$z$ & $H(z)$ & $\sigma_{H(z)}$ & Ref.\\
\hline
0.07 & 69.0 & 19.6 &  \cite{Zhang:2012mp}\\
0.09 & 69 & 12 & \cite{Simon:2004tf}\\
0.12 & 68.6 & 26.2 & \cite{Zhang:2012mp}\\
0.17 & 83 & 8 & \cite{Simon:2004tf}\\
0.179 & 75 & 4 &  \cite{Moresco:2012jh}\\
0.199 & 75 & 5 &  \cite{Moresco:2012jh}\\
0.20 & 72.9 & 29.6 &  \cite{Zhang:2012mp}\\
0.27 & 77 & 14 & \cite{Simon:2004tf}\\
0.28 & 88.8 & 36.6 &  \cite{Zhang:2012mp}\\
0.352 & 83 & 14 & \cite{Moresco:2012jh}\\
0.38 & 83 & 13.5 &  \cite{Moresco:2016mzx}\\
0.4 & 95 & 17 & \cite{Simon:2004tf}\\
0.4004 & 77 & 10.2 & \cite{Moresco:2016mzx}\\
0.425 & 87.1 & 11.2 &  \cite{Moresco:2016mzx}\\
0.445 & 92.8 & 12.9 &  \cite{Moresco:2016mzx}\\
0.47 & 89.0 & 49.6 &  \cite{Ratsimbazafy:2017vga}\\
0.4783 & 80.9 & 9 &  \cite{Moresco:2016mzx}\\
0.48 & 97 & 62 & \cite{Stern:2009ep}\\
0.593 & 104 & 13 &  \cite{Moresco:2012jh}\\
0.68 & 92 & 8 &  \cite{Moresco:2012jh}\\
0.75 & 98.8 & 33.6 &  \cite{Borghi:2021rft}\\
0.781 & 105 & 12 &  \cite{Moresco:2012jh}\\
0.8 & 113.1 &  15.1  &  \cite{Jiao:2022aep}     \\
0.875 & 125 & 17 &  \cite{Moresco:2012jh}\\
0.88 & 90 & 40 &  \cite{Stern:2009ep}\\
0.9 &  117 &  23 &  \cite{Simon:2004tf}\\
1.037 & 154 & 20 &  \cite{Moresco:2012jh}\\
1.3 & 168 & 17 &  \cite{Simon:2004tf}\\
1.363 & 160 & 33.6 &  \cite{Moresco:2015cya}\\
1.43 & 177 & 18 &  \cite{Simon:2004tf}\\
1.53 & 140 & 14 &  \cite{Simon:2004tf}\\
1.75 & 202 & 40 &  \cite{Simon:2004tf}\\
1.965 & 186.5 & 50.4 & \cite{Moresco:2015cya}\\
\hline 
\bottomrule\hline
\end{tabular}
\label{tab:CC}

\end{center}
\end{table}

The chi-squared function $\chi^2_{\rm CC}$ associated with these measurements is formulated as follows:
\begin{equation}
\chi^2_{\rm CC} = \sum_{i=1}^{33} \frac{[H^{\text{obs}}(z_i)-H^{\text{th}}(z_i)]^2}{\sigma^2_{H^{\text{obs}}(z_i)}},
\end{equation}
where $H^{\text{obs}}(z_i)$ represents the observed Hubble parameter value, accompanied by its corresponding standard deviation $\sigma^2_{H^{\text{obs}}(z_i)}$ from the provided table. The counterpart $H^\text{th}(z_i)$ signifies the theoretical Hubble parameter value derived from the cosmological model under consideration.\\

\noindent\textbf{PP and PPSH0ES}: Type Ia supernovae (SNe Ia) have traditionally played a crucial role in establishing the standard cosmological model. These supernovae provide valuable measurements of distance moduli, which in turn constrain the uncalibrated luminosity distance $H_0d_L(z)$ or the rate of late-time expansion. For a supernova at redshift $z$, the theoretical apparent magnitude $m_B$ is defined by the equation:

\begin{eqnarray}
\label{distance_modulus}
m_B = 5 \log_{10} \left[ \frac{d_L(z)}{1\rm Mpc} \right] + 25 + M_B,
\end{eqnarray}
where $M_B$ represents the absolute magnitude. Hence, the distance modulus $\mu(z)$ can be expressed as $\mu(z) = m_{B} - M_{B}$. In the context of a flat cosmology, the luminosity distance is given by: 
\begin{equation}
\label{eq:dl}
d_L(z) = (1+z)\int_0^{z}\frac{dz^\prime}{H(z^\prime)}.
\end{equation}

In this current investigation, we utilize distance modulus measurements of SNe Ia from the Pantheon+ sample \cite{Brout:2022vx}. This dataset consists of 1701 light curves representing 1550 distinct SNe Ia spanning the redshift range $z \in [0.001, 2.26]$, referred to as PP. Additionally, we incorporate SH0ES Cepheid host distance anchors \cite{Brout:2022vx} to our analysis, enabling constraints on both $M_B$ and $H_0$; this dataset is denoted as PPSH0ES.

In the process of constraining cosmological parameters using the PP data, we minimize the following expression:

\begin{equation}
\label{eq:likelihood}
\chi^2_{\rm PP} = \Delta D^T~C_{\rm stat+syst}^{-1}~\Delta D ,
\end{equation}
where $D$ is the vector of 1701 SN distance modulus residuals calculated as
\begin{equation}
\label{eq:dmu}\Delta D_i = \mu_i - \mu_{{\rm model}}(z_i) ,
\end{equation}
comparing the observed supernova distance ($\mu_i$) to the predicted model distance ($\mu_{{\rm model}}(z_i)$) using the measured supernova/host redshift. Here, $C_{\rm stat+syst} = C_{\rm stat} + C_{\rm syst}$, with $C_{\rm stat}$ and $C_{\rm syst}$ denoting statistical and systematic covariance matrices respectively.

Although there exists degeneracy between the parameters $M_B$ and $H_0$ when analyzing SNe alone, we overcome this limitation by incorporating the recently published SH0ES Cepheid host distance anchors (R22) into the likelihood. This modification enables us to place constraints on both $M_B$ and $H_0$. After including SH0ES Cepheid host distances, the SN distance residuals are adjusted as follows:
\begin{equation}
\label{eq:dmuprime}
\Delta D^\prime_i=
        \begin{cases}
            \mu_i - \mu_i^{{\rm Cepheid}} & i \in \text{Cepheid hosts} \\
            \mu_i - \mu_{{\rm model}}(z_i) &\text{otherwise} ,
        \end{cases}
    \end{equation}
where $\mu_i^{{\rm Cepheid}}$ represents the Cepheid-calibrated host-galaxy distance from SH0ES. The incorporation of this information, along with the SH0ES Cepheid host-distance covariance matrix ($C^{\rm Cepheid}_{\rm stat+syst}$) from R22, leads to the modification of the likelihood function:
\[ \chi^2_{\rm PPSH0ES} = \Delta D{^\prime}^T~(C^{\rm SN}_{\rm stat+syst}+C_{\rm stat+syst}^{\rm Cepheid})^{-1}~\Delta D^\prime, \]
where $C^{\rm SN}_{\rm stat+syst}$ refers to the SN covariance. Additional details can be found in the reference \cite{Brout:2022vx}.\\

\noindent\textbf{BBN}: In all of our analyses, we employ a fresh assessment of the physical baryon density denoted as $\omega_b$ (where $\omega_b \equiv \Omega_bh^2$), with a value of $0.02233\pm0.00036$, derived from Big Bang Nucleosynthesis (BBN). This calculation takes into account refined data from experimental nuclear physics, which has been acquired at the Laboratory for Underground Nuclear Astrophysics (LUNA) located at the INFN Laboratori Nazionali del Gran Sasso in Italy ~\cite{Mossa:2020gjc}.

    The baseline free parameters of the anisotropic model, that is, $\Lambda$CDM+$\Omega_{\sigma0 }$, are given by $\mathcal{P}= \left\{ \omega_{\rm b}, \, \omega_{\rm c}, \, H_0, \, \Omega_{\sigma0 }  \right\}$, where $\omega_{\rm b}=\Omega_{\rm b} h^2$ and $\omega_{\rm c}=\Omega_{\rm c}h^2$ are, respectively, the present-day physical density parameters of baryons and CDM, $H_0$ is the Hubble constant and $\Omega_{\sigma0 }$ is the anisotropy expansion parameter. We assume three neutrino species, approximated as two massless states and a single massive neutrino of mass $m_{\nu}=0.06\,\rm eV$. We use uniform priors: $\omega_{\rm b}\in[0.01,0.03]$, $\omega_{\rm c}\in[0.05,0.25]$, $\,H_0\in[60,80]$, $\Omega_{\sigma0 }\in[0,0.001]$, and the publicly available Boltzmann code \texttt{CLASS}~\cite{Blas:2011rf} with the parameter inference code  \texttt{Monte Python}~\cite{Audren:2012wb} to obtain correlated Monte Carlo Markov Chain (MCMC) samples. We analyze the MCMC samples using the python package \texttt{GetDist}\footnote{\href{https://getdist.readthedocs.io/en/latest/intro.html}{https://getdist.readthedocs.io/en/latest/intro.html}}.

\section{Results and Discussion}\label{sec4}
\begin{figure*}[hbt!]
	\centering
	\includegraphics[width=16cm]{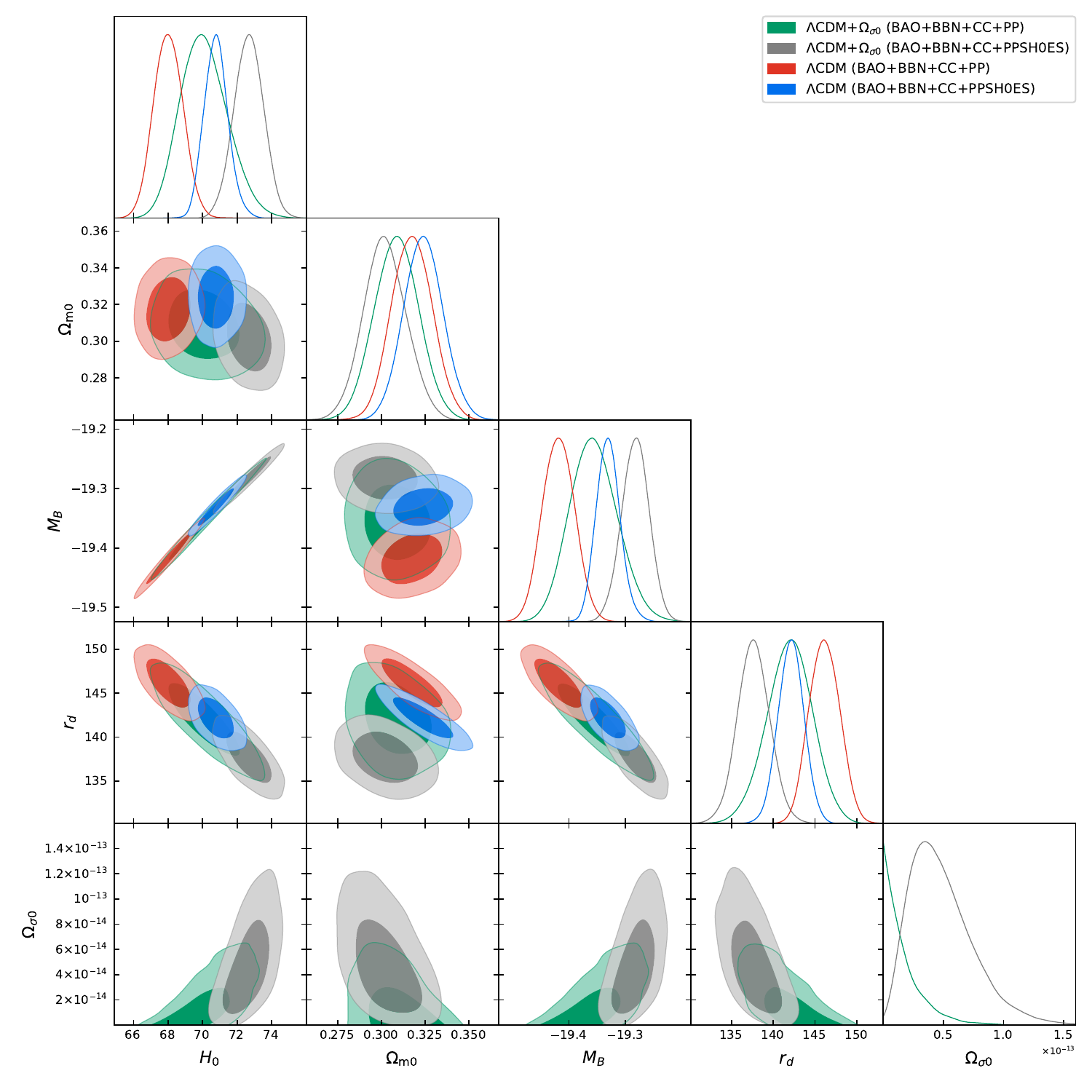}
	\caption{
 One and two dimensional marginalized confidence regions (68\% and 95\% CL) for different parameters of the $\Lambda$CDM and $\Lambda$CDM + $\Omega_{\rm \sigma0}$ models from BAO + BBN + CC + PP and BAO + BBN + CC + PPSH0ES data.
 } 
	\label{sky1}
\end{figure*}

In Table~\ref{tab:1}, we report constraints (68\% CL) on  the free ($\omega_{\rm b}, \, \omega_{\rm c}, \, H_0, \, \Omega_{\sigma0}$) and some derived parameters ($\Omega_{\rm m0}$, $z_{\rm d}$, $r_{\rm d}$) of the $\Lambda$CDM and $\Lambda$CDM+$\Omega_{\sigma0 }$ models from the BAO+BBN+CC+PP and BAO+BBN+CC+PPSH0ES data. In Figure \ref{sky1}, we show one and two dimensional marginalized confidence regions (68\% and 95\%  CL) for different parameters of the anisotropic and $\Lambda$CDM models from BAO+BBN+CC+PP and BAO+BBN+CC+PPSH0ES data. In both cases, the anisotropic parameter $\Omega_{\sigma0 }$ is constrained with the upper bound of the order $10^{-14}$ (95\% CL). In what follows, we analyze the effects of anisotropy on different parameters.

  The constraint on $H_0$ from BAO+BBN+CC+PP data reads: $H_{\rm 0}= 70.1^{+1.1}_{-1.4}$, and in case BAO+BBN+CC+PPSH0ES, it reads:  $H_{\rm 0}= 72.7^{+1.6}_{-1.6}$ which is consistent with SH0ES  measurement: $H_{\rm 0}^{\rm R22}=73.04\pm1.04~{\rm km\, s^{-1}\, Mpc^{-1}}$. From Figure \ref{sky1}, we notice a positive correlation of $H_0$ with $\Omega_{\sigma0}$. So a larger amount of anisotropy would lead to larger value of $H_0$. We note that the mean values of $H_0$ in $\Lambda$CDM+$\Omega_{\rm \sigma0}$ are higher compared to $\Lambda$CDM, and this shift to higher values is not only due to the widening of error bars resulting from the introduction of anisotropy on top of the $\Lambda$CDM ~\cite{Vagnozzi:2019ezj}. Quantifying the $H_0$ tension with SH0ES measurement $H_{\rm 0}^{\rm R22}=73.04\pm1.04~{\rm km\, s^{-1}\, Mpc^{-1}}$ in the two analyses, we find that there are 3.8$\sigma$ and $1.7\sigma$ tensions on $H_0$ in $\Lambda$CDM and $\Lambda$CDM+$\Omega_{\rm \sigma0}$ models respectively with BAO+BBN+CC+PP data. Thus, in this case, the anisotropy reduces the $H_0$ tension by $2.1\sigma$.  In the constraints from BAO+BBN+CC+PPSH0ES data, we find 1.8$\sigma$ tension on $H_0$ in $\Lambda$CDM model but no tension on $H_0$ in $\Lambda$CDM+$\Omega_{\rm \sigma0}$ model. Thus, anisotropy is able to reduce the $H_0$ tension by $\sim 2\sigma$ in both cases.


\begin{table*}[hbt!]
  \caption{Constraints (68\% CL) on  the free and some derived parameters of the $\Lambda$CDM and $\Lambda$CDM+$\Omega_{\sigma0}$ models from BAO+BBN+CC+PP and BAO+BBN+CC+PPSH0ES data. }
  \label{tab:1}
	\scalebox{0.8}{
	\begin{centering}
	  \begin{tabular}{lcccc}
  	\hline\toprule \hline
    \multicolumn{1}{l}{Data set} & \multicolumn{2}{c}{BAO+BBN+CC+PP} & \multicolumn{2}{c}{BAO+BBN+CC+PPSH0ES} \\  \hline
      &$\Lambda$CDM & $\Lambda$CDM+$\Omega_{\sigma0}$ & $\Lambda$CDM  & $\Lambda$CDM+$\Omega_{\sigma0}$    \\\hline 
      
$10^{-2}\omega_{\rm b }$ & $2.233\pm 0.035   $ &  $2.230\pm 0.036$ & $2.258\pm 0.035 $ &   $2.232\pm 0.036   $ 
      \\ 
    
\vspace{0.1cm}
$\omega_{\rm c }$  &$0.1240\pm 0.0068 $ & $0.1288\pm 0.0077 $ & $0.1392\pm 0.0065 $ & $0.1365\pm 0.0065  $ 
  \\ 
 
\vspace{0.10cm} 
$H_{\rm 0}$ [${\rm km\, s^{-1}\, Mpc^{-1}}$]  &    $68.02\pm 0.84  $  & $70.1^{+1.2}_{-1.5} $   & $ 70.79\pm 0.69 $ & $72.67\pm 0.85$\\

\vspace{0.1cm}
$\Omega_{\sigma0 }$ &   0 & $<2.0\times 10^{-14}\;(95\%\; \text{CL})$  & 0 & $<6.5\times 10^{-14}\;(95\%\; \text{CL})$  
   \\

\vspace{0.1cm}
$M_{B}$ [mag]  &   $-19.418\pm 0.028 $ & $-19.356^{+0.039}_{-0.044}$ & $-19.330^{+0.020}_{-0.022}$ & $-19.283\pm 0.024$ \\\hline

\vspace{0.10cm} 
$\Omega_{\rm m0 }$ &   $0.318\pm 0.011 $   & $0.309\pm 0.012 $  & $0.324\pm 0.011 $ & $0.302\pm 0.012  $     \\
\vspace{0.10cm} 
 $z_{\rm d}$ &   $1060.09\pm 0.95 $   & $1060.37\pm 0.98 $  & $1061.75\pm 0.85$ & $1060.99\pm 0.90$     \\
 \vspace{0.10cm} 
 $r_{\rm d}$ &   $146.1\pm 1.8$   & $142.0\pm 2.7$  & $142.1\pm 1.5 $ & $137.7\pm 1.9 $ \\ \hline
\vspace{0.10cm} 
 $\chi^2_{\rm min}$ &   $1438.62$   & $1438.58$  & $1334.4$ & $1322.94$ \\
\hline\bottomrule\hline
\end{tabular}
\end{centering}}
\end{table*}

From Figure \ref{sky1}, we notice that $\Omega_{\sigma0}$ and $\Omega_{\rm m0 }$ are negatively correlated in both cases, that means, larger $\Omega_{\sigma0}$ values correspond to smaller $\Omega_{\rm m0}$ values. $\Omega_{\rm m0}$  is constrained to  $\Omega_{\rm m0 }=0.318\pm 0.011\; (0.324\pm 0.01) $ and   $\Omega_{\rm m0 }=0.309\pm 0.012\; (0.302\pm 0.012)$ from the BAO+BBN+CC+PP (BAO+BBN+CC+PPSH0ES) data in the $\Lambda$CDM and $\Lambda$CDM+$\Omega_{\rm \sigma0}$ models, respectively.  In $\Lambda$CDM+$\Omega_{\rm \sigma0}$, we find a shift of $0.6\sigma$ ($1.4\sigma$) to lower $\Omega_{\rm m0}$-values from the  BAO+BBN+CC+PP (BAO+BBN+CC+PPSH0ES) data  compared to $\Lambda$CDM.    \\

Next, from Figure \ref{sky1}, we notice that $\Omega_{\sigma0}$ and $M_B$ are positively correlated in both cases, that means, larger $\Omega_{\sigma0}$ values correspond to larger $M_B$ values. $M_B$  is constrained to  
$M_B=-19.418\pm 0.028\;(-19.330^{+0.020}_{-0.022})$ and $M_B=-19.356^{+0.039}_{-0.044}\; (-19.283\pm 0.024)$  from the BAO+BBN+CC+PP (BAO+BBN+CC+PPSH0ES) data in the $\Lambda$CDM and $\Lambda$CDM+$\Omega_{\rm \sigma0}$ models, respectively.  In $\Lambda$CDM+$\Omega_{\rm \sigma0}$, we find a shift of $1.2\sigma$ ($1.5\sigma$) to higher $M_B$-values from the  BAO+BBN+CC+PP (BAO+BBN+CC+PPSH0ES) data  compared to $\Lambda$CDM.    \\

Further, from Figure \ref{sky1}, we notice that $\Omega_{\sigma0}$ and $r_{\rm d}$ are negatively correlated in both cases, that means, larger $\Omega_{\sigma0}$ values correspond to smaller $r_{\rm d}$ values. $r_{\rm d}$  is constrained to
$r_{\rm d}=146.1\pm 1.8\;(142.1\pm 1.5 $ and   $r_{\rm d}=142.0\pm 2.7\;(137.7\pm 1.9) $  from the BAO+BBN+CC+PP (BAO+BBN+CC+PPSH0ES) data in the $\Lambda$CDM and $\Lambda$CDM+$\Omega_{\rm \sigma0}$ models, respectively.  In $\Lambda$CDM+$\Omega_{\rm \sigma0}$, we find a shift of $1.3\sigma$ ($1.8\sigma$) to lower $r_{\rm d}$-values from the  BAO+BBN+CC+PP (BAO+BBN+CC+PPSH0ES) data  compared to $\Lambda$CDM.    \\



From the above observations, we deduce that different parameters of the model are influenced due to anisotropy. Also, in the process of multi-parameter simultaneous fitting, parameters influence each other \cite{Marra:2021fvf,Camarena:2023rsd}. For instance, one may notice strong positive correlation between $H_0$ and $M_B$ in Figure \ref{sky1}. Nonetheless, the presence of anisotropy affects the parameters of $\Lambda$CDM in such a way that the $H_0$ tension is relaxed by $\sim 2\sigma$ in both the analyses under consideration.  Further, we emphasize that in the previous studies \cite{Akarsu:2019pwn,Akarsu:2021max} of the anisotrpic model under consideration, fixed values of $z_{\rm d}$ were utilized in the BAO analyses, but here $z_{\rm d}$ is allowed to vary. The main results in these studies include the analyses with CMB data but here we have obtained CMB independent constraints on the anisotropic model, and that too with some updated and different data combinations compared to \cite{Akarsu:2019pwn,Akarsu:2021max}.   


The observation that anisotropy could lead to a reduction in $r_{\rm d}$ and an increase in $H_0$ is a remarkable insight. The anisotopic model behaves like the Early Dark Energy model (EDE) wherein $a^{-6}$ scaling holds relevance in the early Universe, and it becomes negligible at later times (see the recent review \cite{Poulin:2023lkg} on EDE models, and references therein).  In fact, the resolutions using EDE lead to a reduction in the sound horizon's scale by postulating the presence of supplementary energy density prior to recombination. As a result, this augmentation of energy density contributes to an elevation in the Hubble parameter during the period surrounding photon decoupling. This allows us to reinterpret the success of EDE within the framework of an anisotropic model. However, the implications when the CMB is introduced need further elucidation. It is important to note that attempting anisotropic cosmology using equation \eqref{themodel} is not sufficient as it lacks the necessary inclusion of angular considerations on the celestial sphere. Yet, our analysis provides a promising direction. There is mounting evidence suggesting that the Universe may deviate from the assumptions of FLRW model, as suggested by ~\cite{Aluri:2022hzs}. Several potential directions of anisotropy have emerged as well. Notably, the CMB exhibits a distinct hemispherical power asymmetry, favoring a phenomenological Bianchi model over the $\Lambda$CDM model, as indicated in the review ~\cite{Aluri:2022hzs}. Intriguingly, significant shifts in $H_0$ are observed along the axis of this anomaly when the CMB is masked, as detailed in ~\cite{Yeung:2022smn}. This is unlikely to be a coincidental finding. Therefore, with a careful reinstatement of angular considerations and meticulous modeling of the CMB, a positive outcome could be attained. Anisotropy is also evident in SNe Ia datasets, which indicate a preference for a higher $H_0$ in the direction of the CMB dipole in the CMB frame, as demonstrated by references such as ~\cite{Krishnan:2021jmh} (Pantheon), ~\cite{Zhai:2022zif} (SH0ES+Pantheon), and ~\cite{McConville:2023xav} (Pantheon+). The latter two studies illustrate that the combination of Cepheids and Supernova data yields substantial variations in $H_0$. Moreover, there are bulk flows aligned with the Shapley supercluster's direction that deviate from the expectations of the $\Lambda$CDM model, as discussed in ~\cite{Watkins:2023rll,Whitford:2023oww}.  Particularly intriguing is the finding that an anisotropy can emulate the effects of EDE and influence the BAO scale. 

\section{Conclusion}
\label{sec5}
In this work, we have reported CMB independent constraints on $H_0$ in an ansotropic model, namely $\Lambda$CDM+$\Omega_{\sigma0}$, using the BBN, BAO, CC, PP and PPSH0ES data. In the anisotropic model, we have obtained $H_{\rm 0}=70.1^{+1.2}_{-1.5}\; (72.67\pm 0.85)\;\rm km\, s^{-1}\, Mpc^{-1}$ both with 68\% CL from BAO+BBN+CC+PP (BAO+BBN+CC+PPSH0ES) data.  The analyses of the anisotropic model with the two combinations of data sets reveal that anisotropy is positively correlated with $H_0$, and an anisotropy of the order $10^{-14}$ in the anisotropic model reduces the $H_0$ tension by $\sim 2\sigma$. Overall, the study provides insights into how an anisotropic extension of the standard cosmological model, combined with different cosmic probes, can influence the estimation of the Hubble constant and potentially contribute to resolving the existing tension in its measured value.

\begin{acknowledgements}
The author gratefully thanks the reviewers for fruitful comments and suggestions to enhance the quality of the paper, and also acknowledges Eoin O Colgain, Eleonora Di Valentino, Sunny Vagnozzi and Suresh Kumar for useful discussions and valuable suggestions to improve the paper. 
\end{acknowledgements}
\bibliography{refs}
\end{document}